\documentclass[times,twocolumn,final]{elsarticle}

\usepackage{medima}
\usepackage{framed,multirow}

\usepackage{amssymb}
\usepackage{latexsym}

\usepackage{url}
\usepackage{xcolor}
\usepackage{booktabs}
\usepackage[bookmarks=false]{hyperref}
\usepackage{comment}
\usepackage{multirow}
\usepackage{tablefootnote}

\definecolor{newcolor}{rgb}{.8,.349,.1}

\journal{Under review}

\begin{document}

\verso{Diedre Carmo \textit{et~al.}}

\begin{frontmatter}

\title{MEDPSeg: Hierarchical polymorphic multitask learning for the segmentation of ground-glass opacities, consolidation, and pulmonary structures on computed tomography}


\author[1]{Diedre S. \snm{Carmo}\corref{cor1}}
\cortext[cor1]{Corresponding author: d211492@dac.unicamp.br}
\author[1]{Jean A. \snm{Ribeiro}}
\author[2]{Alejandro P. \snm{Comellas}}
\author[3]{Joseph M. \snm{Reinhardt}}
\author[3]{Sarah E. \snm{Gerard}}

\author[1]{Letícia \snm{Rittner}}
\author[1]{Roberto A. \snm{Lotufo}}

\address[1]{School of Electrical and Computer Engineering, Universidade Estadual de Campinas, Campinas, 13083-852, Brazil}
\address[2]{Department of Internal Medicine, University of Iowa, Iowa City, 52242, USA}
\address[3]{Roy J. Carver Department of Biomedical Engineering, University of Iowa, Iowa City, 52242, USA}

\received{22 March 2024}

\begin{abstract}
The COVID-19 pandemic response highlighted the potential of deep learning methods in facilitating the diagnosis, prognosis and understanding of lung diseases through automated segmentation of pulmonary structures and lesions in chest computed tomography (CT). Automated separation of lung lesion into ground-glass opacity (GGO) and consolidation is hindered due to the labor-intensive and subjective nature of this task, resulting in scarce availability of ground truth for supervised learning. To tackle this problem, we propose MEDPSeg. MEDPSeg learns from heterogeneous chest CT targets through hierarchical polymorphic multitask learning (HPML). HPML explores the hierarchical nature of GGO and consolidation, lung lesions, and the lungs, with further benefits achieved through multitasking airway and pulmonary artery segmentation. Over 6000 volumetric CT scans from different partially labeled sources were used for training and testing. Experiments show PML enabling new state-of-the-art performance for GGO and consolidation segmentation tasks. In addition, MEDPSeg simultaneously performs segmentation of the lung parenchyma, airways, pulmonary artery, and lung lesions, all in a single forward prediction, with performance comparable to state-of-the-art methods specialized in each of those targets. Finally, we provide an open-source implementation with a graphical user interface at \url{https://github.com/MICLab-Unicamp/medpseg}.
\end{abstract}


\end{frontmatter}

\section{Introduction}

The COVID-19 pandemic showcased the possible contributions of automated deep learning segmentation methods to assist medical doctors in the diagnosis and prognostication of lung diseases using chest computed tomography (CT) segmentation and analysis~\citep{shi2020review, carmo2021rapidly, shan2021abnormal, bhosale2023application}. These methods also contribute to the understanding of pathophysiological processes of novel diseases through medical research, such as correlating lung parenchyma lesions in CT with Long COVID demographic information~\citep{carmo2023evidence}. The development and improvement of automated segmentation methods for pulmonary lesions and structures in CT images is an important step in enabling rapid and more accurate CT analysis in the future, establishing deep learning-based methods as producers of a reliable and effective image-based biomarker for medical decision support.


\begin{figure}[ht]
\centerline{\includegraphics[width=\columnwidth]{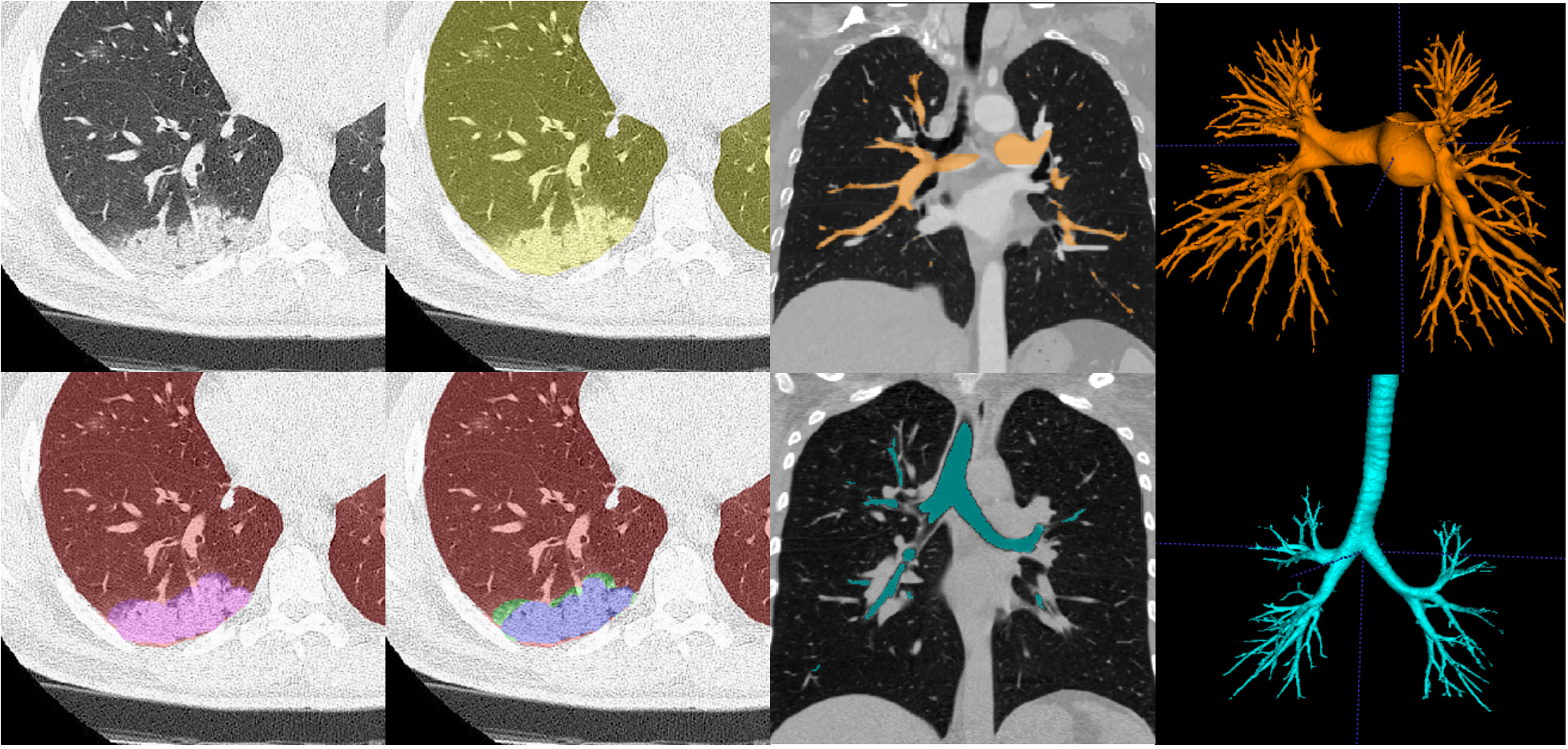}}
\caption{\label{fig:intro}Illustration of all target formats used in this work. Axial slices show the hierarchical properties of lung lesion annotation with whole lung in yellow, healthy tissue in red, lesions in magenta, GGO in green, and consolidated lung in blue. The airways and the pulmonary arterial tree are displayed in cyan and orange, respectively. All visualizations created with ITK-Snap~\citep{yushkevich2006user}.}
\end{figure}

Quantitative analysis of CT images relies on accurate segmentation of regions of interest, such as identifying regions of lung inflammation to assess pneumonia severity~\citep{shan2021abnormal}. However, manual segmentation of pulmonary structures in CT is prohibitively time-consuming and cost-intensive. Increased availability of training data, recent advancements in deep learning, and UNet-like convolutional neural networks (CNN)~\citep{ronneberger2015u, isensee2021nnu} have accelerated performance improvement in the automated segmentation of chest CT over the last decade~\citep{carmo2022systematic}. Nevertheless, the performance and generalization capabilities of deep learning methods are still relatively poor in tasks with a limited amount of high-quality manual annotation, such as the task of separation of lung lesions into ground-glass opacities (GGO) and parenchymal consolidation~\citep{fan2020inf, yang2023mmvit}. In this work, we propose a way to circumvent this data variability limitation through hierarchical polymorphic multitask learning (HPML). Hierarchical polymorphic learning explores the hierarchical properties of lung annotations, indirectly learning complex GGO and consolidation labels using simpler general lung lesion or lung parenchyma annotations. In parallel, multitask learning further enhances the proposed method capabilities through the inclusion of segmentation heads for anatomically related chest CT targets in the form of the pulmonary artery and the airway tree (Fig.~\ref{fig:intro}). The proposed methodology named Modified EfficientDet for Polymorphic Pulmonary Segmentation (MEDPSeg) achieves comparable performance to state-of-the-art specialized methods while efficiently multitasking pulmonary artery, airway, lung parenchyma, and lung lesion chest CT targets. In addition, we show PML enables MEDPSeg to establish a new state-of-the-art for GGO and consolidation segmentation.

\subsection{Related Work}

For segmentation of pulmonary targets in chest CT, recent work has achieved stable and generalizable results for lung parenchyma, even in the presence of abnormalities. \cite{hofmanninger2020automatic} suggests that the large amount of data variability used in their method, Lungmask, was the most important contribution for its impressive performance. nnUNet~\cite{isensee2021nnu}, an adaptable framework that automatically trains an UNet model to the provided imaging and label pairs, makes heavy use of data preprocessing, normalization, and augmentation to achieve state-of-the-art performance in multiple medical imaging segmentation challenges, also raising the importance of the data processing pipeline step and data variance. Challenges that nnUNet was in the top performing submissions include the Airway Tree Modelling challenge (ATM 2022)~\citep{zhang2023multi}, Pulmonary Artery Segmentation Challenge (PARSE 2022)~\citep{luo2023efficient} and the COVID-19 Lung CT Lesion Segmentation Challenge~\citep{roth2022rapid}. The most commonly studied lung lesions are lung cancer nodules and opacities~\citep{carmo2022systematic}. Regarding opacities, recent medical studies still use approximations of lung involvement for disease severity quantification~\citep{cho2022quantitative}, without precise voxel-wise segmentation, using methods such as AMFM~\citep{xu2006mdct}. Nevertheless, there have been many proposals for automated voxel-wise segmentation of COVID-19 pneumonia-related lesions~\citep{shi2020review}. CopleNet~\citep{wang2020noise} uses a noise-robust Dice~\citep{sudre2017generalised} loss modification and an exponential moving average of a novel architecture as a teacher network that is only updated when the student has a low loss. \cite{app11125438} refines a Lungmask~\citep{hofmanninger2020automatic} lung segmentation into opacity areas through unsupervised vessel artifacts exclusion and k-means clustering. Biondi et al.'s justification for not using supervised or semi-supervised deep learning stems from the low availability and variability of available opacity manual annotation. This annotated data variability is a key factor in the generalization and performance for deep learning-based chest CT segmentation. However, the more complex the annotation is, one will find less available gold standard manually annotated data~\citep{carmo2022systematic}. Although data augmentation partially solves this issue~\citep{isensee2021nnu}, having as many CT images with manually annotated targets from as many sources as possible is still desirable~\citep{hofmanninger2020automatic}. 

The voxel-wise separation of general lung lesions into specific GGO and consolidation regions is an example of a chest CT segmentation task with a scarce amount of available data for supervised training. It is a very subjective task with high interrater variability even before lesion separation~\citep{sotoudeh2022multi}. One common approach used to distinguish GGO and consolidated lung is to apply a simple HU intensity threshold~\citep{lu2021quantitative}. Even when more involved techniques are used, evaluation is commonly performed through quantification measurements or severity scores~\citep{chaganti2020automated, cho2022quantitative}. This uncertainty of segmentation originates mainly from the undefined boundary between opacity and healthy tissue~\citep{lensink2022soft}, and is made even more prevalent when also having to define the border between consolidation and opacity~\citep{silva2010illustrated}. The SemiSeg dataset~\citep{fan2020inf, MedSeg2021} is one of the only available resources for pixel-wise GGO and consolidation manual annotation, with 100 annotated slices. Yet, many works using SemiSeg choose to abstain from this challenging task and only use it for general lung lesion segmentation~\citep{zhang2021exploiting, qiu2021miniseg, rao2023covid, zhan2024easwin}. The original SemiSeg publication tried to circumvent the limited data availability problem through semi-supervised learning, leveraging information from unlabeled data~\citep{fan2020inf}. Their proposed method InfNet uses a random propagation strategy that generates pseudo labels for unlabeled data using a trained network, and iteratively trains new networks with these pseudo labels. There are many other examples of GGO and Consolidation segmentation methods working with the SemiSeg dataset. COVID TV-UNet~\citep{saeedizadeh2021covid} imposes connectivity regularization in its UNet loss function to improve baseline UNet results. Adversarial semi-supervised learning also has achieved success in segmentation of GGO and consolidation, by regularizing pseudo labels with variational auto encoders and discriminators together with the limited supervised learning examples~\citep{jin2022efficient}. In some cases, applications of famous segmentation architectures also led to competitive results, such as DeepLabV3+~\citep{polat2022multi}. A lightweight transformer architecture, MMVITSeg~\citep{yang2023mmvit}, achieved impressive performance by merging encoder-decoder design with small visual transformers while keeping the number of parameters orders of magnitude lower than similarly performing methodologies. UNet-PSGR~\citep{jia2023convolutional} is an enhanced UNet architecture with pixel-wise sparse graph reasoning, a hybrid CNN and Graph Neural Network approach to improve the modeling of long-range dependencies through the graph-based representation. 

Another way to introduce more broad knowledge to the network is through multitask learning. A limitation of deep learning-based medical image segmentation methods is the adherence to a specific label format from specific datasets, due to the inherent limitation of a fixed number of output channels in CNN architecture design. Usually, the involvement of multiple structures includes a single one-hot annotation for all structures in all training cases~\citep{henschel2020fastsurfer}. Alternatively, using multiple networks or multiple segmentation heads allows different label formats and partially labeled datasets by using multiple decoder branches with a different number of output channels~\citep{carmo2020multiattunet, tan2021sgnet, zhang2021dodnet}. However, this approach is not suitable in applications with a scarce amount of manual annotations available for specific tasks, e.g., separating general lung lesions into specific GGO and consolidation regions. The network or segmentation head dedicated to the low data availability task would be undertrained. DoDNet~\citep{zhang2021dodnet} proposed a new way, through conditioning. The same network produces different outputs conditioned by a prompt. This is also becoming more and more common in other fields of machine learning, with large pre-trained models being able to be conditioned through prompts for multitasking. One disadvantage of this approach, however, is the computational efficiency of such large models, sometimes not being able to be run in consumer-grade computers, and the large pretraining requirements~\citep{zhou2023comprehensive}.

Multitasking can also result in neighboring anatomy guidance in medical imaging segmentation. For example, unsupervised segmentation of the rib cage and airways can be used to constrain the lung parenchyma location and to guide lung lobe segmentation~\citep{carmo2022systematic}. Another form of anatomy guidance is leveraging the hierarchical properties of the labeled structures. This approach has shown promising results both in natural~\citep{li2022deep} and medical~\citep{bakas2018identifying} images, however, with annotations still confined to the same dataset. \cite{gerard2021ct} proposed polymorphic learning as a way to accommodate datasets with heterogeneous lung labels and exploit hierarchical relationships in the lung. In their work, polymorphic learning was used to separate the lung into its lobes in COVID-19 patients without using any training data from this domain, and instead utilizing whole lung labels from existing datasets of multiple species with only nonspecific whole lung labels.


Given this literature context, we believe that the separation of lung lesions into more specific labels of GGO and consolidation can be improved, by utilizing hierarchical polymorphic multitask learning to accommodate diverse datasets with heterogeneous hierarchically or anatomically related segmentation targets. 

\begin{table*}[ht]
\centering

\resizebox{\textwidth}{!}{
\begin{tabular}{lllll}
\toprule
\textbf{Dataset Name} & \textbf{Label Format} & \textbf{N. images (split)} & \textbf{Slices/scan} & \textbf{Voxel size (mm³)}                                                           \\\midrule
\multicolumn{5}{l}{\textbf{Gold Standard Data}} \\\midrule
IdeiaGov~\citep{carmo2021rapidly}   & Lesion  & 119 (95/12/12) & 335$\pm$82 &0.60$\pm$0.32    \\
MOSMED~\citep{morozov2020mosmeddata} & Lesion  & 50 (40/5/5) & 41$\pm$4& 4.29$\pm$1.22          \\
MICCAI-C~\citep{roth2021rapid} & Lesion & 199 (149/20/20) & 69$\pm$37 & 3.03$\pm$0.90            \\
CoronaCases~\citep{ma_jun_2020_3757476} & Lesion  & 10 (-/-/10) & 258$\pm$39 & 0.59$\pm$0.15 \\
MSC~\citep{MedSeg2021} & Separation & 9 (7/1/1) & 92$\pm$116 & 2.44$\pm$0.76            \\
SemiSeg~\citep{fan2020inf} & Separation & 100 (45/5/50)  & N/A & N/A\\
LongCI (in-house) & Separation & 90 (-/-/90)  & N/A & N/A\\
ATM~\citep{zhang2023multi} & Airway & 399 (339/30/30) & 663$\pm$140 & 0.33$\pm$0.19\\
Parse~\citep{luo2023efficient} & Vessel & 100 (80/10/10) & 303$\pm$29 & 0.46$\pm$0.11\\\midrule
\multicolumn{5}{l}{\textbf{Silver Standard Data}} \\\midrule
IdeiaGov~\citep{carmo2021rapidly}& All auto. & 1148 & 399$\pm$100 & 0.54$\pm$0.25 \\
MOSMED~\citep{morozov2020mosmeddata}& All auto. & 848 & 42$\pm$4 & 4.41$\pm$1.00 \\
Luna16 (LIDC-IDRI)~\citep{setio2017validation} & All auto.  & 800 & 256$\pm$134 & 0.78$\pm$0.44 \\
MSD (Lung)~\citep{antonelli2022medical} & All auto. & 50 & 280$\pm$110 & 0.83$\pm$0.30 \\
Zaffino COVID-19~\citep{zaffino2021open}& All auto. & 64 & 802$\pm$185 & 0.20$\pm$0.05  \\
TCIA Covid 19~\citep{an2020ct}& All auto.  & 457  & 71$\pm$46 & 2.98$\pm$0.89   \\
NSCLC-Radiomics~\citep{aerts2014decoding} & All auto. & 259 & 124$\pm$22 & 2.86$\pm$0.02        \\
COVID-19-AR~\citep{desai2020chest}& All auto. & 19 & 552$\pm$201 & 0.84$\pm$0.62  \\
COVID-19-NY-SBU~\citep{saltz2021stony}& All auto.  & 478  & 228$\pm$233 & 1.64$\pm$1.03                 \\
COPD-TLC~\citep{regan2011genetic} & All auto. & 365 & 563$\pm$66 & 0.20$\pm$0.04 \\\midrule
\multicolumn{2}{l}{\textbf{Total}}             & 6150 & \multicolumn{2}{l}{}   \\\bottomrule                                              
\end{tabular}
}
\caption{\label{tab:datadesc}Summary of gold and silver standard data sources with their respective number of images and train/val/test splits, mean and standard deviation (std) of axial slices per scan, voxel size (mean$\pm$std) as computed by SimpleITK~\citep{beare2018image}. Label format refers to what structures are annotated to what level of detail. Lesion annotation has healthy and unhealthy area labels, separation has healthy, GGO and consolidation areas annotated, and airway and vessel are binary airway and pulmonary artery annotations. LongCI is an in-house dataset, will be made publicly available soon. Silver standard data has automated annotations of all targets, and gold standard refers to manual annotation.}
\end{table*}
\subsection{Contributions}

The proposed HPML methodology advances the state-of-the-art in GGO and consolidation separation by taking advantage of the hierarchical properties of lung anatomy, i.e, each voxel in an image can be labeled as lung vs. non-lung; the lung can be more specifically labeled as healthy lung parenchyma vs lung lesion; lung lesion can be more specifically labeled as GGO vs. consolidation. HPML indirectly optimizes the complex GGO and consolidation target using a polymorphic head. Neighboring anatomy guidance through simultaneous learning of airway and pulmonary artery in the multitask head is shown to improve lesion separation performance. Multitask performance on the additional targets of lung parenchyma, airway, and pulmonary artery is shown to be statistically equivalent or superior to specialized state-of-the-art methods through experiments involving more than 6000 volumetric CT scans from multiple partially labeled sources. By using a single backbone encoder and the polymorphic multihead strategy, all structures can be predicted in a single model inference. Additionally, the model is lightweight and can be run on consumer grade hardware, and we contribute open source code with a graphical user interface for reproducibility.

The contributions of this work can be summarized as follows:

\begin{itemize}
\item Hierarchical polymorphic multitask learning (HPML) is shown to enable state-of-the-art performance in GGO and consolidation segmentation.
\item The unified framework proposed in this work provides competitive results when compared to specialized state-of-the-art methods in pulmonary artery, airway, lung parenchyma, and lung lesion segmentation, while also being computationally efficient. This is the first method to multitask the aforementioned chest CT pulmonary targets.
\item MEDPSeg's results and evaluation are easily reproducible in consumer-grade hardware, with open-source code, public trained weights, Python pip package, and software tool with a graphical user interface. 
\end{itemize}


\section{Data}
\label{sec:data}
Multiple independent data sources were used in this study. Datasets with voxel-level annotation performed by human experts are referred to as ``gold standard" datasets (Section~\ref{sec:gold}). Datasets annotated with automated methods are referred to as ``silver standard" (Section~\ref{sec:silver}). Note that no dataset includes all labels, i.e., we learn from partially labeled datasets.

\subsection{Gold standard data}
\label{sec:gold}
Each CT image with manual voxel-level annotations contains one of the following targets: healthy and lesion areas; GGO and consolidation; airway; or pulmonary artery (Table~\ref{tab:datadesc}).  

IdeiaGov, MOSMED, and MICCAI-C datasets are combined into a development cohort named Lesion representing lesion-level annotations, which was randomly split into 80/10/10 percent train/validation/testing sets. The combined dataset is highly heterogeneous representing a wide variety of voxel sizes, manufacturers, acquisition parameters, and subtle differences in annotation protocol. SemiSeg, a 2D slice-wise dataset, is separated following the original publication 45/5/50 (\%) holdout split~\citep{fan2020inf} and combined with MSC 80/10/10 (\%) split into the Separation cohort. LongCI is a recently developed in-house dataset of Long COVID opacity and consolidation manual annotation performed by the consensus of three blind expert pulmonologist raters. At the time of writing, it is being submitted for publication and open access. It is kept as an external lesion separation dataset, isolated from development. CoronaCases is also kept as an external test dataset, but for general lesion segmentation. For datasets containing only lesion annotation, such as MICCAI-C, MOSMED and SemiSeg, lung masks were included with the application of lungmask~\citep{hofmanninger2020automatic} followed by visual inspection of results. Finally, ATM and Parse represent airway and pulmonary artery annotations being used in the multitask branch of MEDSeg and were also split into 80/10/10 (\%) holdout splits. For more details on acquisition parameters, please refer to the cited references shown in Table~\ref{tab:datadesc}. All gold standard datasets are anonymized and were collected outside of the scope of this project, being made available publicly or as part of medical imaging segmentation challenges.

\subsection{Silver standard data}
\label{sec:silver}
Datasets without manual annotation come from a variety of sources with different acquisition parameters and scanners, and were originally collected to study diseases such as chronic obstructive pulmonary disease (COPD), lung cancer, and COVID-19. Note that the subset of MOSMED without annotation is different from the annotated subset in the gold data (Tab.~\ref{tab:datadesc}). Some of these data sources were stored in DICOM format with multiple series and modalities. All silver data were used only for training. Silver standard data contributes mainly with whole lung parenchyma labels for polymorphic learning, acquired through the application of Lungmask~\citep{hofmanninger2020automatic}.

An additional way we used silver standard data is through silver standard pretraining (SSP). Labels for all target formats were acquired from multiple teacher nnUNet. These teacher nnUNets~\citep{isensee2021nnu} are trained in our gold standard data for lesion (L-nnUNet), airway (A-nnUNet) and pulmonary artery (PA-nnUNet). For the segmentation of GGO and consolidation, the L-nnUNet lesion area is approximately separated using a -300 HU threshold~\citep{lu2021quantitative}. With all these labels, the network is initially trained only in silver standard data. Subsequent training continues as normal with silver standard data providing only lung annotation examples. All datasets were publicly available except for COPD-TLC, approved under the University of Iowa Institutional Review Board (IRB), consisting of acquisitions of the COPDGene project~\citep{regan2011genetic} acquired at the University of Iowa.


\subsection{Data Curation}
Not every single one of the data contained in the mentioned data sources is useful for this research. For gold standard data, scans from CoronaCases with "radiopaedia" in the name were removed due to wrong headers. For the COPD-TLC data from the University of Iowa, only total lung capacity (TLC) scans were included. For silver standard data, some of the datasets were made available in their raw DICOM format and hierarchy without any curation. DICOM hierarchy is composed of multiple studies and series per patient, with multiple series pertaining to the same acquisition date, representing test scans or just orientation changes and other requests from the physician. To avoid repeated scans, we developed a simple heuristic to select a single DICOM series per study. The following filters are applied until only one "Original, Primary, Axial" ImageType series is left.

\begin{itemize}
\item Remove any series with ImageType that is not "Original, Primary, Axial" or "Original, Primary, Axial, Helix".
\item Select the softer ConvolutionKernel, following the per Manufacturer list provided by~\citep{mackin2019matching}.
\item If there are still multiple series, return the one with more axial slices.
\end{itemize}



\section{Methodology}
\label{sec:method}
\begin{figure*}[ht]
\centerline{\includegraphics[width=\textwidth]{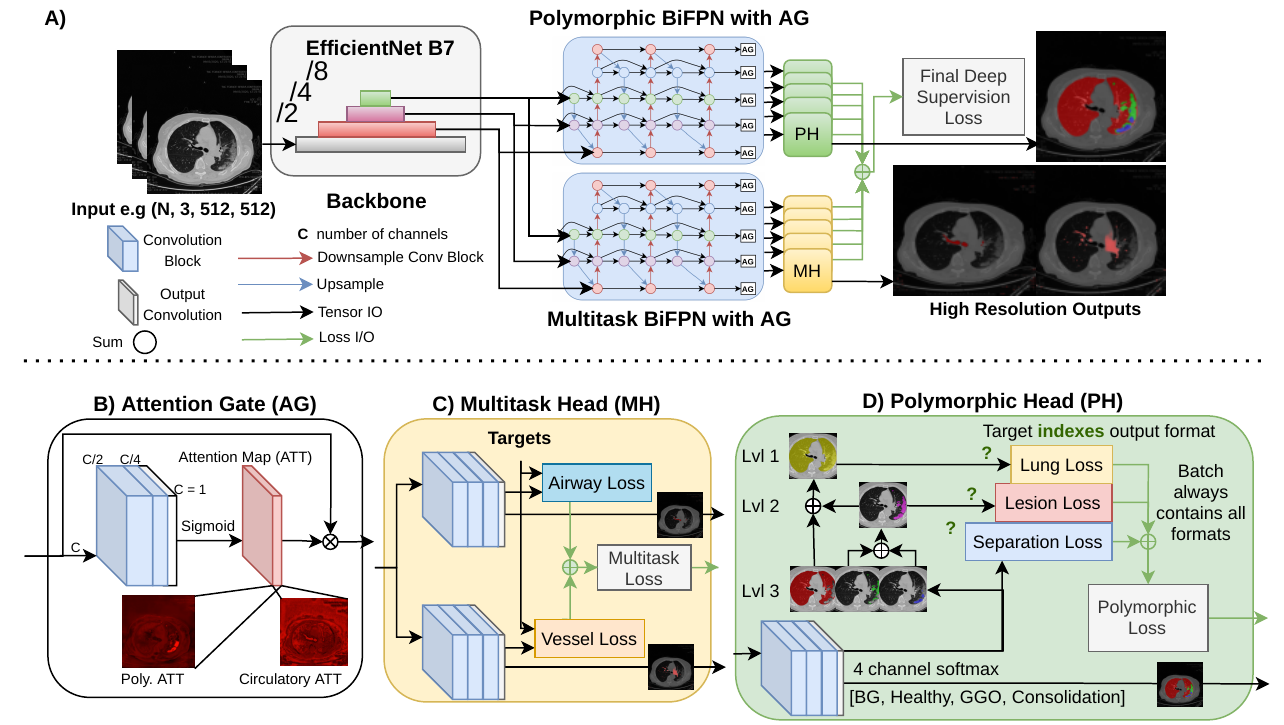}}
\caption{\label{fig:arch}(A) Illustration of MEDPSeg architecture from initial feature extraction with the EfficientNet backbone, BiFPN with spatial attention (B) feature decoding, and the multitask (C) and polymorphic (D) head outputs. Multiple output formats are built on the fly with sum reductions and the applicable format is applied depending on the target format of the input image.}
\end{figure*}

In this section, we summarize how the full methodology of the proposed Modified EfficientDet for Polymorphic Pulmonary Segmentation (MEDPSeg) works (Fig.~\ref{fig:arch}). Starting with data preparation steps (Section~\ref{sec:dataprep}), description of the low level architecture (Section~\ref{sec:arch}), details of how the architecture is optimized in multiple partially labeled datasets with HPML (Sections~\ref{sec:details}), training hyperparameters (Section~\ref{sec:hyper}) and finally our evaluation methodology (Section~\ref{sec:eval}).

\subsection{Data preparation and augmentation}
\label{sec:dataprep}
\color{black}The heterogeneity of the data, not only in acquisition parameters but also in label format, requires standardization. After careful consideration of different normalization and preprocessing techniques with early experiments, all CT images were clipped to the range $[-1024, 600]$ Hounsfield unit (HU), and then normalized to the range $[0, 1]$. This window contains enough information to contrast both the lung and mediastinal structures in pulmonary CT images. Silver data was cropped to the lung region using Lungmask's~\citep{hofmanninger2020automatic} lung segmentation. For gold data, only slices with annotation are used in training, a strategy that led to more balanced label representation and performance in early experiments. In evaluation, predictions are made on all slices of the input CT volume (Fig.~\ref{fig:inference}).

MEDPSeg, leverages information from all target formats and datasets, but recall that all data sources are partially labeled. Therefore, a standard representation is necessary for heterogeneous target encodings. We need to have all annotations on the same batch, but technical limitations of tensors being stored as fixed shape arrays require a fixed number of channels. Therefore, a 4 channel representation is used for the whole batch. Each item in a batch can have one of the following target formats: Lung (background: channel 0, lung: channel 1); Lesion (background: 0, healthy lung: 1, lesion: 2); Lesion Separation (background: 0, healthy lung: 1, GGO: 2, consolidation: 3); Airway (background: 0, airway 1); and finally Pulmonary Artery (background: 0, vessel: 1). Given the different sizes of each dataset, a custom data sampler was implemented to sample and augment each annotation format randomly and ensure every batch has representation from every target format, with 30000 samples per format (lung, lesion, separation, airway, artery) for a total of 150000 samples per epoch. Note that lung samples always come from silver standard data. Each sample is accompanied by metadata indicating the target format. When a target format has less than 30000 annotated samples, such as GGO and consolidation separation or pulmonary artery, we still perform a uniform random sampling, repeating samples. To alleviate this repetition problem, data augmentation is performed on the fly during training with random scaling and rotation, mirroring of all axis, Gaussian noise, Gaussian blur, brightness and contrast intensity augmentation, and finally a $256x256$ random crop. For the target and only for lesion labels, we perform random morphology to try and counter the noisy nature of these annotations. For all cited augmentations we follow the same parameters as Isensee et al.'s 2D nnUNet, sourced from their supplementary material~\citep{isensee2021nnu}. In silver standard pretraining (Section~\ref{sec:silver}), all targets annotated by automated methods are represented in a 6-channel format (background: 0, healthy lung: 1, GGO: 2, consolidation: 3, vessel: 4, airway 5). 

\subsection{Architecture}
\label{sec:arch}
\begin{figure}[ht]
\centerline{\includegraphics[width=\columnwidth]{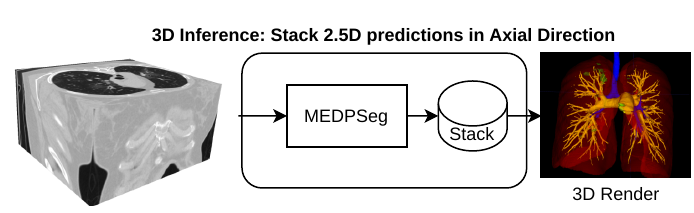}}
\caption{\label{fig:inference}During validation and for inference, MEDPSeg processes whole volumes through stacking axial predictions.}
\end{figure}

Modified EfficientDet for Polymorphic Pulmonary Segmentation (MEDPSeg) works with 2.5D input slices, where each batch sample has 3 axial slices composed of the target slice and its immediate neighbors, regardless of the original spacing. This representation provides local volumetric and slice thickness information. Note, however, in cases where there is no volumetric information (e. g. SemiSeg GGO and consolidation separation dataset~\citep{fan2020inf}) the same slice is repeated in the three channels. Therefore, MEDPSeg can be applied to both 3D and 2D contexts, with MEDPSeg output for a given volumetric CT scan being generated by stacking the results of 2.5D predictions in the axial direction for each slice (Fig.~\ref{fig:inference}). Earlier non-PML versions of this architecture have been shown to provide superior performance in lung opacity segmentation when compared to traditional 3D UNets~\citep{carmo2021multitasking}.

As indicated by the name, Modified EfficientDet for Polymorphic Pulmonary Segmentation is inspired by EfficientDet's usage of a large backbone EfficientNet~\citep{tan2019efficientnet} network and feature pyramid networks (FPN)~\citep{tan2020efficientdet}. Tan et al. arrived at an optimized FPN named Bidirecational FPN (BiFPN). BiFPN outputs feature maps of varying spatial-resolution with a fixed number of output channels. These feature maps can be combined and used for any downstream task, such as classification, object detection, or pixel-level segmentation. MEDPSeg uses high-resolution backbone features (starting from /2 the original input resolution to /8) from an ImageNet pre-trained EfficientNet-B7~\citep{tan2019efficientnet}. The EfficientNet input stem convolution is replaced with a 3D 3x3 convolution with no padding to take advantage of the 2.5D Input. In a single forward pass, these same backbone features are shared and processed by two 128-channel BiFPN. One of the BiFPNs is specialized in the polymorphic segmentation of the lung. This BiFPN will build features toward the optimization of the hierarchical segmentation of the lung, lung lesion, GGO and consolidation. The other BiFPN is specialized in airway and pulmonary artery segmentation, and therefore building features specialized for the segmentation of tubular tree-like pulmonary structures for the circulation of air or blood. 

These BiFPNs are modified with spatial attention gates~\citep{gorriz_assessing_2019} at the outputs (BiFPN-AG, Fig.~\ref{fig:arch}B), to spatially filter and enhance features at each spatial resolution of the feature pyramid. Spatial attention gates also have the benefit of providing explainability to where the network is ``looking at". In BiFPN-AG, each spatially weighted output feature is linearly upsampled (x2), resulting in five multi-scale feature maps from input resolution to /16 the original resolution.  The polymorphic BiFPN-AG has polymorphic segmentation heads (PH) with a softmax output of four fixed channels. These channels are reduced through sum operations depending on the level of detail of the current target, with GGO plus consolidation representing lesioned lung, and lesioned plus healthy lung representing the whole lung (Fig~\ref{fig:arch}D). The multitask BiFPN-AG has multitasking segmentation heads (CH) with two softmax outputs of two channels each, for multitasking both airway and pulmonary artery (Fig~\ref{fig:arch}C). 

In the segmentation heads (CH or PH), features are processed by three convolution blocks with separable convolutions~\citep{chollet2017xception}, batch normalization~\citep{ioffe_batch_2015} and swish~\citep{ramachandran2017searching}. The key to making MEDPSeg's architecture work is how to optimize polymorphic and multitask training using PH and CH outputs through a customized loss function. The loss will analyze each target format separately and combine each result into a mean final loss value as the starting point for backward propagation. The following section will go through this process from top to bottom.

\subsection{Hierarchical Polymorphic Multitask Learning}
\label{sec:details}
\begin{figure}[ht]
\centerline{\includegraphics[width=\columnwidth]{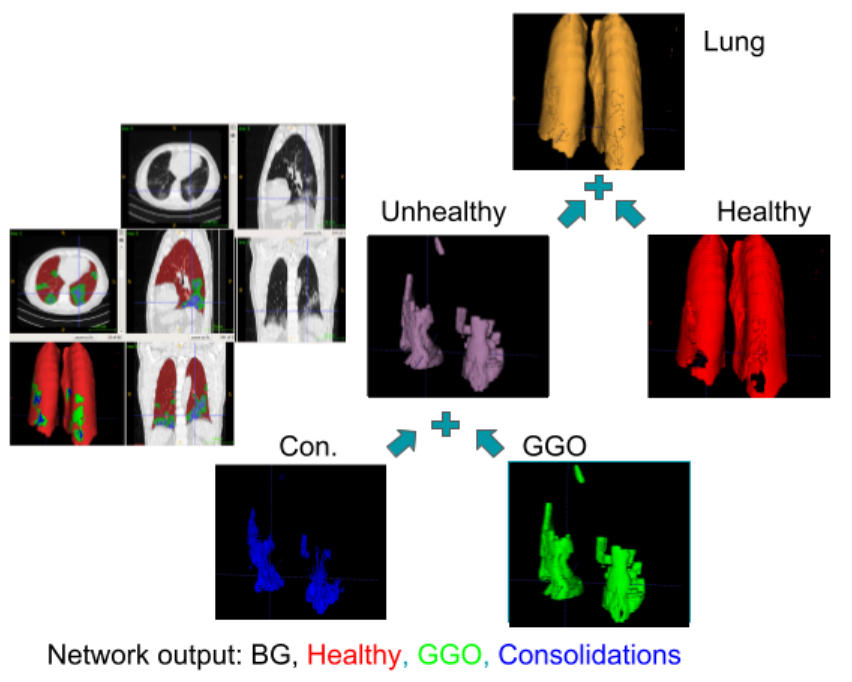}}
\caption{\label{fig:graph_summary}Hierarchical Polymorphic Learning takes advantage of the hierarchical natures of annotations of lungs and pathologies. Airway and pulmonary artery labels are multitasked in separate outputs due to not being hierarchically consistent with the pathology hierarchy tree.}
\end{figure}

The properly prepared and augmented batch of 2.5D axial slices goes through many steps for the HPML optimization of the architecture. HPML is in summary a way to leverage information from multiple partially labeled datasets for the optimization of a fixed architecture. For the polymorphic head, depending on the target format, the four-channel output (background, healthy lung, GGO, and consolidation) is reduced through sums to match the target format. This process is referred to as hierarchical polymorphic learning (HPL) (Fig~\ref{fig:graph_summary}), i.e, match how much of the target is annotated. For the multitask head segmenting the airway and pulmonary artery, the loss is simply computed directly on the airway and pulmonary artery outputs, when the target is present.

Careful consideration is taken such that for every batch, the loss components for each target format are computed separately and summed together. The polymorphic head contributes to the lung loss, lesion loss and separation loss components. Depending on the target format, the fixed four channel output (background, healthy lung, GGO and consolidation) is reduced through sums accordingly to match the target format (Fig~\ref{fig:arch}D). This is possible due to the mathematical property that the derivative of a sum of functions is the sum of the individual derivatives. This allows automatic differentiation to work even with polymorphic learning. With this approach, we indirectly propagate gradients to the optimization of all four output channels even when the target doesn't contain that information directly, but a suitable prediction can be built by summing the relevant outputs together. For example, a lesion prediction can be built by summing the GGO and consolidation outputs, and the whole lung can be built by summing the lesion with the healthy lung output. This approach allows us to merge large datasets with more readily available simpler labels, with smaller datasets with more complex manual labels while always optimizing the four output channels. The multitask head directly contributes with airway and vessel outputs and losses. The final $Loss$ for optimization is computed by aggregating the contribution from all target formats: 

\begin{equation}
\label{eq:loss_simple}
TotalLoss = \frac{L_{airway} + L_{vessel} + L_{lung} + L_{lesion} + L_{separation}}{5}
\end{equation}

The definition of a proper loss function $L$ for the generation of gradients towards the optimization of correct segmentation is still necessary. For every component of $Loss$, let the underlying loss function be $L$, with 0-indexed channel $c$ of a total of $C$ channels and corresponding softmax output segmentation $\hat{y_{c}}$ and target segmentation $y_{c}$. $L$ is the sum of generalized dice loss (GDL)~\citep{sudre2017generalised}, without background, and cross entropy loss with background: $L = GDL(\hat{y}, y) + CrossEntropyLoss(\hat{y}, y)$. These components can be defined separately as follows:

\begin{equation}
GDL(\hat{y_{c}}, y_{c}) = 1 - 2\frac{\sum_{c=1}^{C-1}w_{c}\sum_{}^{}\hat{y_{c}}y_{c}}{\sum_{c=1}^{C-1}w_{c}\sum_{}^{}(\hat{y_{c}}+y_{c})}
\end{equation}

\begin{equation}
CrossEntropyLoss(\hat{y_{c}}, y_{c}) = -\sum_{c=0}^{C-1}log(\hat{y_{c}})y_{c}
\end{equation}

In GDL, a channel contribution is inversely proportional to its area, reducing the effects of class imbalance in small targets. On the other hand, cross-entropy penalizes errors more heavily, and in this implementation, includes background optimization. The intent is to leverage advantages from both approaches, as in the common combination of Dice Loss and Cross Entropy used by other works~\citep{isensee2021nnu}.

Notice that output features from the BiFPN-AG in Figure~\ref{fig:arch}A are from input resolution to one-sixteenth the input resolution, with multiple PH and CH in different spatial resolutions. The multiple resolution PH and CH outputs are used for deep supervision~\citep{wang2015training}, where the loss function is applied in places other than the main high resolution output to inject gradients into deeper weights more effectively. The contribution to the deep supervision loss $DSL$ from the main high-resolution loss $L_{0}$ and each of the four lower resolution head loss functions $L_{i}$ with $i\in[1, 4]$ is weighted in the following way:

\begin{equation}
DSL = 0.75L_{0} + \sum_{i=1}^{4}2^{-(i+2)}L_{i}
\end{equation}

Therefore, 75\% of the weight is on the optimization of the main output, and the other 25\% is distributed to the optimization of the lower resolution outputs with exponential decay. The $Loss$ for optimization is computed by aggregating the contribution from all heterogeneous target formats, in the same way as equation~\ref{eq:loss_simple} but replacing $L$ with $DSL$. In other words, the average of $DSL$ applied to all of the five target formats separately.

\subsubsection{Architectural Variations}
\label{sec:var}
For clarity, some naming conventions have to be established to avoid confusion among variations of this final MEDPSeg architecture. In some cases for the remainder of this manuscript, the architecture is specialized on a single target format with a single BiFPN, such as training only for lesion segmentation or only for airway segmentation. There is no multitasking or polymorphic head. This target specialized variation also doesn't employ deep supervision and spatial attention at the BiFPN. This architectural variation will be named Specialized Modified EfficientDet for Segmentation (S-MEDSeg).

\begin{figure}[ht]
\centerline{\includegraphics[width=\columnwidth]{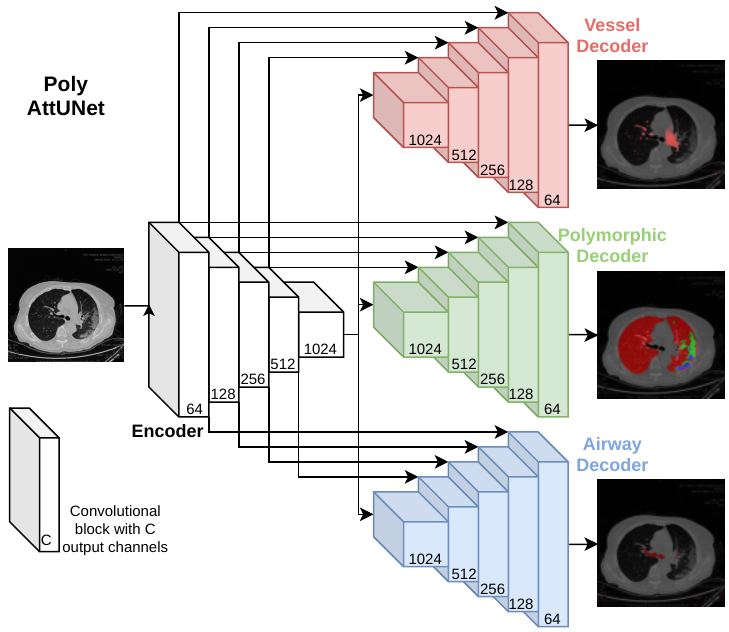}}
\caption{\label{fig:poly_att_unet}Illustration of Poly AttUNet (P-AttUNet). A simpler UNet-like approach to HPML with one encoder and three decoders.}
\end{figure}

Finally, to explore the impact of HPML in the more common UNet architecture, we also explored a parallel approach using a relatively simpler 2.5D AttUNet~\citep{carmo2020multiattunet} architecture, with a polymorphic variation P-AttUNet (Fig~\ref{fig:poly_att_unet}) designed by simply having features from a large encoder (1024 channels at bottleneck) used by a main polymorphic decoder and two additional airway and vessel decoders.

\subsection{Training Hyperparameters}
\label{sec:hyper}
 
\begin{figure}[ht]
\centerline{\includegraphics[width=\columnwidth]{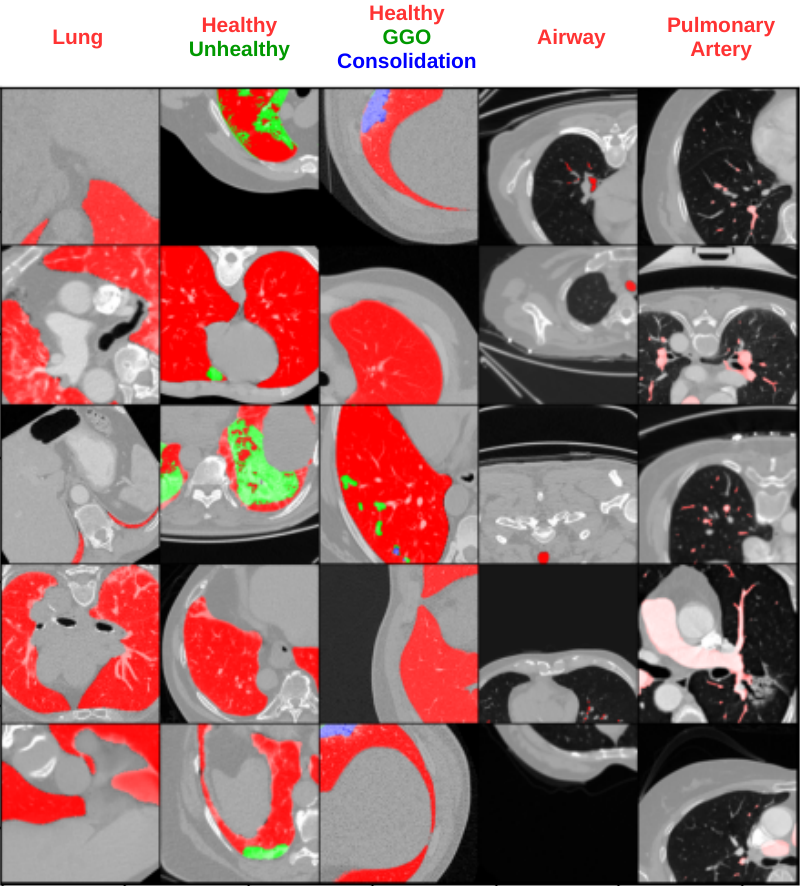}}
\caption{\label{fig:batch_indexing}A representative batch of data used during training. 5 samples from each of the 5 label formats are displayed organized by columnn. RGB encoded targets are overlapped with the corresponding CT axial slice.}
\end{figure}

Some fixed hyperparameters determined in early experiments are AdamW as the optimizer with initial learning rate $1e-4$ and weight decay $1e-5$; exponential learning rate decay of $0.985$; the mean of cross entropy and generalized dice loss~\cite{sudre_generalised_2017} for the loss function; and, for gold standard training, batch size of $25$ (Fig~\ref{fig:batch_indexing}). Possible lower values while keeping a balanced representation of all targets would be multiples of 5: 1 from each target (5); 2 from each target (10); and so on. Note that degraded performance is observed with lower batch sizes. The best weight is selected by the lower validation loss over 3D scans, with a volumetric output being built on-the-fly by stacking predictions (Fig~\ref{fig:inference}). We select the best validation loss weight after 120 hours (approximately 50 epochs when including all data) of training in a A100 80GB GPU or 3080Ti 12 GB GPU, depending on the memory requirements of the experiment. It is important to highlight that the trained MEDPSeg uses less than 8GB of VRAM during inference, and doesn't require expensive hardware for reproducibility of its segmentation capabilities through the provided tool~\footnote{\url{https://github.com/MICLab-Unicamp/medpseg}}. Silver standard pre-training uses batch size 1 of one 512x512 fully (automatically) annotated slice. 

\subsection{Evaluation}
\label{sec:eval}

To evaluate the validation performance during training and final testing performance, adequate quantitative metrics are computed with 3D targets and 3D predictions built by stacking axial predictions. The following quantitative evaluation metrics are employed in this work: firstly, Dice~\citep{sudre2017generalised}, false positive error (FPE) and false negative error (FNE) are used as implemented by SimpleITK~\citep{beare2018image}. FNE represents the rate at which positive voxels were classified as negative, and is also equal to $1 - Sensitivity$. FPE represents the rate of positive predictions that were false positives. Higher FPE and FNE represent worse performance. Note that SimpleITK's definition of FPE does not include the number of negative voxels and subsequently has no bias to increased field of view or smaller segmentations, a problem that happens with specificity and other implementations of FPE. For comparisons with other methods in the SemiSeg dataset, evaluation follows the methodology proposed by the original publication: slice-wise Dice, specificity and sensitivity~\citep{fan2020inf}, using publicly provided implementations from the ATM22 challenge~\citep{zhang2023multi}.

Dice is used as a general overlap metric that takes into account both false positives and false negatives. Note that Dice tends to assume a higher value in large structures, such as for the lung parenchyma, and the larger the structure, the less perceptible the penalization due to small errors~\citep{maier2022metrics}. In addition, since lesions are not always present, a special case needs to be established as follows: when no target or prediction is present, the Dice is assumed to be 1. Mathematically, this can be represented by adding a small value $\epsilon$ in the numerator and denominator. When a metric is by definition not defined, such as sensitivity without positive cases, that specific case is not considered for that metric computation. 

These overlap and voxel-wise metrics are not sufficient for evaluating tubular segmentation tasks, and we adopt the tree length detected rate (TD) and branch detected rate (BD) of the largest connected component, as implemented by the evaluation pipeline from the ATM22 airway segmentation challenge~\citep{zhang2023multi}. The main challenges regarding the segmentation of tubular tree-like structures can be summarized in two main phenomena: leakage and breakage. Leakage refers to segmented areas going beyond the tubular borders due to unclear boundaries and breakage refers to results containing disconnected branches. Both the airway and pulmonary artery are singular connected structures where automated segmentations can suffer from these problems. These topological problems are not detected well by overlap metrics such as Dice that give more weight to the larger structures close to the tree root. TD is defined as the ratio between the total detected tree length and the total length of the airway tree in the ground truth.

\begin{equation}
TD = \frac{TL_{det}}{TL_{ref}}
\end{equation}

\begin{equation}
BD = \frac{BN_{det}}{BN_{ref}}
\end{equation}

Where $TL_{det}$ is the total length of the prediction and $TL_{ref}$ is the total length of the ground truth. Similarly, BD denotes the ratio between the number of detected branches $BD_{det}$ and the number of branches in the ground truth $BD_{ref}$. Skeletonization and branch counting processes necessary to arrive at these metrics were computed with the implementation provided by the ATM22 challenge, originally made available by Zheng et al.~\citep{zheng2021alleviating}\footnote{\url{https://github.com/haozheng-sjtu/3d-airway-segmentation}}.

The structure of evaluation for the remainder of this manuscript will have one of the following objectives: studying the impact in quantitative metrics of key design choices and hyperparameters in test splits of each involved target format; or comparing the multitasking results of MEDPSeg to specialized methods. Regarding statistical testing, whenever results are compared in terms of significance, the Wilcoxon signed-rank test with $p < 0.05$ is used~\citep{carmo2023automated}. Finally, qualitative evaluations are also performed by examining 3D renders of the worst and best cases according to ground truth.

\section{Experiments and Results}


\label{sec:results}
In this section, we report the results of experiments performed to assess the performance of MEDPSeg. Firstly, our main breakthrough in GGO and consolidation segmentation performance when compared to state-of-the-art methods is presented in (Section~\ref{sec:ggo}), with additional experiments showcasing the impact of HPML and other design choices. Secondly (Section~\ref{sec:multitasking}), we examine the multitasking performance when compared to baseline nnUNets and other methods from the literature in each of the additional targets, also presenting the impact of HPML in multitasking performance. In addition, reproducibility, computational efficiency (Section~\ref{sec:rep}) and qualitative results (Section~\ref{sec:quali}) are explored. For all experiments, unless noted otherwise, the parameters detailed in Section~\ref{sec:method} are used.

\begin{table*}[ht]
\resizebox{\textwidth}{!}{
\begin{tabular}{cccccccc}
\toprule
\multicolumn{3}{c}{\textbf{Parameters}}                             & \multicolumn{2}{c}{\textbf{\begin{tabular}[c]{@{}c@{}}SemiSeg \\ (Test)\end{tabular}}}                                                 & \textbf{\begin{tabular}[c]{@{}c@{}}Lesion Cohort \\ (Test)\end{tabular}} & \multicolumn{2}{c}{\textbf{LongCI}}                                                                                                                                            \\\midrule
\textbf{Architecture}                    & \textbf{HPL} & \textbf{M} & \textbf{\begin{tabular}[c]{@{}c@{}}GGO \\ (2D Dice)\end{tabular}} & \textbf{\begin{tabular}[c]{@{}c@{}}Con. \\ (2D Dice)\end{tabular}} & \textbf{\begin{tabular}[c]{@{}c@{}}Lesion\\ (Dice)\end{tabular}}             & \multicolumn{1}{c}{\textbf{\begin{tabular}[c]{@{}c@{}}GGO \\ (2D Dice)\end{tabular}}} & \multicolumn{1}{c}{\textbf{\begin{tabular}[c]{@{}c@{}}Con. \\ (2D Dice)\end{tabular}}} \\\midrule
AttUNet~\citep{carmo2020multiattunet}     &             &            & $0.58\pm0.21$  & $0.46\pm0.29$ & $0.37\pm0.24$     &  $0.40\pm0.21$  &     $0.23\pm0.27$ \\
P-AttUNet                                & \checkmark  & \checkmark & $0.63\pm0.21$  & $\mathbf{0.57\pm0.28}$ & $\mathbf{0.63\pm0.20}$  &    $\mathbf{0.62\pm0.15}$  &  $\mathbf{0.40\pm0.37}$    \\\midrule
B4 S-MEDSeg                              &             &            & $0.57\pm0.23$  & $0.44\pm0.29$ & $0.24\pm0.18$  &    $0.17\pm0.16$      &    $0.37\pm0.47$   \\
B7 S-MEDSeg                              &             &            & $0.66\pm0.16$ & $0.50\pm0.29$ & $0.46\pm0.21$ &   $0.57\pm0.18$  &   $0.32\pm0.37$  \\
MEDPSeg                                  & \checkmark  &            & $\mathbf{0.63\pm0.20}$ & $0.55\pm0.29$  & $0.63\pm0.22$ &  $0.53\pm0.15$   &   $0.32\pm0.37$   \\
MEDPSeg                                  & \checkmark  & \checkmark & $\mathbf{0.65\pm0.20}$ & $\mathbf{0.58\pm0.29}$  & $\mathbf{0.66\pm0.20}$  &   $\mathbf{0.57\pm0.18}$   &    $\mathbf{0.42\pm0.40}$ \\\bottomrule
\end{tabular}
}
\caption{\label{tab:sizeloss}Experiments in GGO and Consolidation segmentation on the SemiSeg test set, Lesion cohort test set and the LongCI unseen dataset. Our UNet implementation in the form of AttUNet serves as baselines and is enhanced by HPML. Finally, S-MEDSeg's performance is enhanced by increased EfficientNet backbone size (B4 to B7), followed by polymorphic learning (HPL) and airway and pulmonary artery multitasking (M) improvements. Top two results in bold.}
\end{table*}
\subsection{GGO and consolidation separation}
\label{sec:ggo}
Firstly, we explore training, architectural, and general hyperparameter variations. Initial experiments use only scarce gold data GGO and consolidation labels from the Separation cohort. Experiments involving hierarchical polymorphic learning (HPL) introduce Lesion cohort training data learned through polymorphic heads. Experiments with pulmonary artery and airway multitasking (M) also include training data from these targets and the accompanying multitask segmentation head. Testing uses the SemiSeg testing split and the LongCI dataset as an external unseen test set. Models are also tested in a reduction to the Lesion cohort test split. To test in lesion data, the GGO and consolidation outputs are reduced with a sum operation into a general lesion label. These experiments studied the impact of backbone size, architecture variation, and HPML (Tab.~\ref{tab:sizeloss}).

The benefits of increasing backbone size are apparent. A larger backbone from EfficientNet B4 to B7~\citep{tan2019efficientnet} improved performance. Turning specialized S-MEDSeg into multitasking (M) and polymorphic learning (PL) MEDPSeg brought noticeable, significant improvements, including on generalization to general lesion labels and the unseen LongCI data. Performance improvements from PL and M were also noted when augmenting AttUNet into P-AttUNet. The difficulty of consolidation segmentation in LongCI is notable, most likely due to the lower severity of these cases. To assess GGO and consolidation performance when compared to other methods specifically designed for this problem, we compare our best MEDPSeg results with results reported by the literature in the public and well-defined SemiSeg test split (Tab.~\ref{tab:semiseg_compare}).

\begin{table}[ht]
\centering
\resizebox{\columnwidth}{!}{
\begin{tabular}{ccccccc}
\toprule
\textbf{Dataset}     & \multicolumn{6}{c}{\textbf{SemiSeg}}                                                                                                             \\\midrule
\textbf{Target}      & \multicolumn{3}{c}{\textbf{GGO}}                                                 & \multicolumn{3}{c}{\textbf{Consolidation}}                  \\\midrule
\textbf{Method}                           & \textbf{Dice}  & \textbf{Sen.} & \textbf{Spec.}  & \textbf{Dice}  & \textbf{Sen.} & \textbf{Spec.}  \\\midrule
\cite{fan2020inf}                  & $0.62$         & $0.62$               & $0.97$                & $0.46$         & $0.51$               & $0.97$                \\
\cite{isensee2021nnu}           & $\mathbf{0.65}$         & $0.60$               & $0.99$       & $0.50$         & $0.46$               & $0.99$                \\
\cite{saeedizadeh2021covid}       & $\mathbf{0.65}$         & $\mathbf{0.76}$      & $0.98$                & $0.54$& $\mathbf{0.56}$               & $0.99$                \\
\cite{jin2022efficient}        & $0.59$         & $0.61$               & $0.96$                & $0.32$         & $0.33$               & $0.76$                \\
\cite{polat2022multi} & 0.60  &  0.70  &  0.97 & 0.51  & 0.59  & 0.99 \\    
\cite{yang2023mmvit}             & $0.63$         & $0.66$               & $0.96$                & $0.35$         & $0.36$               & $0.70$                \\
\cite{jia2023convolutional}     & $0.62$         & $0.69$               & $0.98$                & $0.49$         & $0.65$      & $0.98$                \\

\multicolumn{1}{c}{MEDPSeg (this work)}             & $\mathbf{0.65}$& $0.64$               & $\mathbf{0.99}$       & $\mathbf{0.58}$& $0.54$               & $\mathbf{0.99}$       \\\bottomrule               
\end{tabular}
}
\caption{\label{tab:semiseg_compare}Our best MEDPSeg compared on the SemiSeg test dataset with slice-wise mean metrics Dice, Sensitivity (Sen.) and Specificity (Spec.) as reported by other methods from the literature. Polat reports 5-fold results instead of using Fan et al.'s testing split. Top results in bold.}

\end{table}

When looking at the challenging problem of GGO and consolidation separation, with limited availability of annotated data, the proposed method outperforms multiple recent works achieving state-of-the-art GGO and consolidation performance in the SemiSeg test dataset, with significantly better Dice on consolidation. Experiments have shown this results directly from specific design choices, most importantly, the network architecture, multitasking with the multitask and polymorphic heads, and polymorphic learning.

\begin{table*}[ht]
\centering
\resizebox{\textwidth}{!}{
\begin{tabular}{ccccccccc}
\toprule
\textbf{Method}    & \textbf{HPL}                             & \textbf{M}                              & \textbf{SSP} & \textbf{Airway (Dice)} & \textbf{Vessel (Dice)} & \textbf{Lesion (Dice)} & \textbf{GGO (2D Dice)} & \textbf{Con. (2D Dice)} \\\midrule
S-MEDSeg  &                                &                                &              & $\mathbf{0.92\pm0.04}$ & N/A                    & N/A                    & N/A                    & N/A                     \\
S-MEDSeg  &                                &                                &              & N/A                    & $\mathbf{0.85\pm0.03}$ & N/A                    & N/A                    & N/A                     \\
S-MEDSeg  &                                &                                &              & N/A                    & N/A                    & $\mathbf{0.66\pm0.21}$ & N/A                    & N/A                     \\
S-MEDSeg  &                                &                                &              & N/A                    & N/A                    & $0.46\pm0.21$          & $\mathbf{0.66\pm0.16}$ & $0.50\pm0.29$           \\\midrule
MEDPSeg   & \checkmark &                                &              & N/A                    & N/A                    & $0.63\pm0.22$          & $0.57\pm0.27$          & $0.55\pm0.29$           \\
P-AttUNet &     \checkmark      & \checkmark &              & $0.88\pm0.05$          & $0.84\pm0.03$          & $0.63\pm0.20$          & $0.63\pm0.21$          & $0.57\pm0.28$           \\
MEDPSeg   & \checkmark & \checkmark &              & $\mathbf{0.90\pm0.04}$ & $0.84\pm0.04$          & $0.64\pm0.22$          & $0.61\pm0.23$          & $\mathbf{0.58\pm0.31}$  \\
MEDPSeg   & \checkmark & \checkmark & \checkmark   & $0.90\pm0.05$          & $\mathbf{0.88\pm0.02}$ & $\mathbf{0.66\pm0.20}$ & $\mathbf{0.65\pm0.20}$ & $\mathbf{0.58\pm0.29}$  \\ \bottomrule
\end{tabular}
}
\caption{Polymorphic multitask learning experiments showcasing the impact of hierarchical polymorphic learning (PL) and multitasking (M) separetely in MEDPSeg and AttUNet when compared to baselines specialized in each target format. SSP refers to silver standard pretraining. N/A means not applicable. Top two results in bold.}
\label{tab:poly}
\end{table*}
\subsection{Multitasking} 
\label{sec:multitasking}
The polymorphic and multitask heads bring not only improvements to GGO and consolidation segmentation performance but also turn MEDPSeg into a more complete lung assessment method, delivering multiple pulmonary targets. An ablation study was performed to evaluate the contribution of components of the proposed polymorphic model (MEDPSeg), when compared to specialized S-MEDSeg trained for each target. Different experiments varied some key design choices such as airway and pulmonary artery segmentation tasks. In addition, the effects of using silver standard data as a pretraining step (SSP) (Tab.~\ref{tab:poly}). 

Results show that the inclusion of pulmonary structure targets (airway and pulmonary artery) is essential for the best performance of MEDPSeg in lesions, GGO, and consolidation. It is worth noting that the performance of silver standard trained MEDPSeg is poor when evaluated in gold standard test labels. However, using it as weight initialization, i.e., silver standard pre-training (SSP), resulted in the best overall performance when looking at all targets at the same time. Although not included in tables, data augmentation, the usage of deep supervision, and spatial self-attention were also essential for achieving the best performance. Note that the convergence of MEDPSeg is relatively difficult. We were not able to achieve competitive results in all targets while using a 3D Polymorphic AttUNet or 3D MEDPSeg variation. In addition, there was a need to separate the vessel and airway outputs on the multitask head for best results. A single three-channel softmax had poor performance in comparison to two, two-channel, softmax outputs. This is most likely due to partially labeled datasets and the fact we have no gold standard annotation of both airway and vessel on the same scan.

\begin{figure}[ht]
\centerline{\includegraphics[width=\columnwidth]{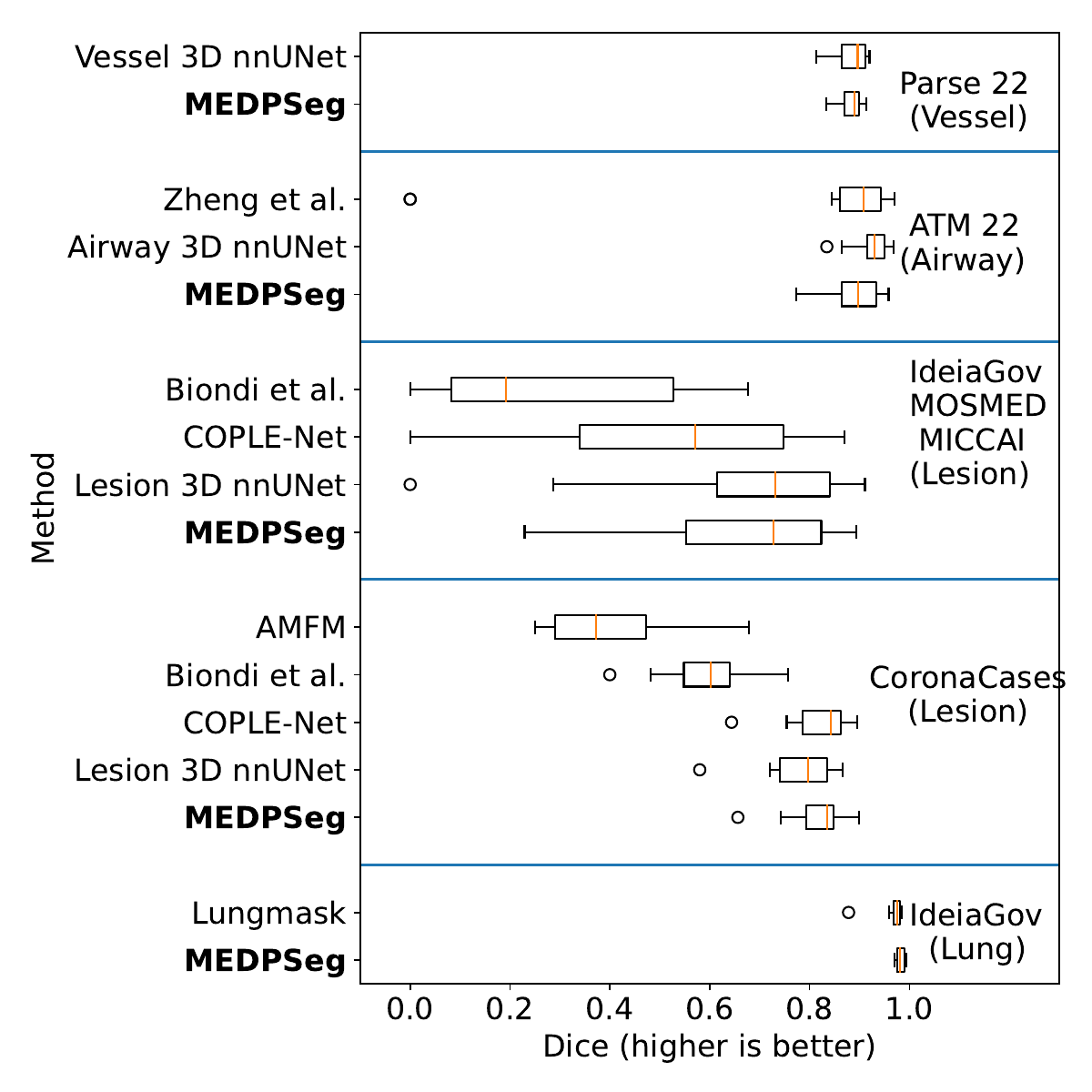}}
\caption{\label{fig:boxplots}Boxplots of Dice with the same best MEDPSeg in bold compared to multiple state-of-the-art methods in different targets.}
\end{figure}

The next experiment compares the performance of our best MEDPSeg model against other methods from the literature which are specialized in each of the targets MEDPSeg outputs. Comparisons are mainly done with specialized 3D nnUNets (2021). This choice of baseline is due to two main reasons: firstly, the ease of training with the same data splits we use for MEDPSeg of the very established nnUNet framework~\footnote{\url{https://github.com/MIC-DKFZ/nnUNet}}. Secondly, 3D nnUNet represents a significantly challenging baseline, given its recent top ranking in medical imaging segmentation challenges~\citep{isensee2021nnu, antonelli2022medical}. For each target format, a nnUNet instance is trained with the same training and validation splits as MEDPSeg. MEDPSeg is also evaluated against Lungmask in lung parenchyma segmentation in the IdeiaGov test data, containing severe pneumonia cases with a median lung involvement of approximately 40\%. For the included methods that are not nnUNet: COPLE-Net~\citep{wang2020noise}, \cite{app11125438}, Lungmask~\citep{hofmanninger2020automatic}, AMFM~\citep{xu2006mdct}, and~\cite{zheng2021alleviating}, we reproduced their prediction capabilities with the provided trained weights and/or implementation code and used the same evaluation implementation for all methods. Note that these methods used different training data. The boxplot in Figure~\ref{fig:boxplots} showcases a general idea of multitasking segmentation performance through Dice. All MEDPSeg results are from the same trained weights. 

\begin{table*}[ht]

\resizebox{\textwidth}{!}{
\begin{tabular}{lcccccccc}
\toprule
\multicolumn{1}{c}{\multirow{2}{*}{\textbf{Method}}} & \multicolumn{2}{c}{\textbf{\begin{tabular}[c]{@{}c@{}}Vessel \\ (PARSE)\end{tabular}}} & \multicolumn{2}{c}{\textbf{\begin{tabular}[c]{@{}c@{}}Airway\\ (ATM)\end{tabular}}} & \multicolumn{2}{c}{\textbf{\begin{tabular}[c]{@{}c@{}}Lesion\\ (CoronaCases)\end{tabular}}} & \multicolumn{2}{c}{\textbf{\begin{tabular}[c]{@{}c@{}}Lung\\ (IdeiaGov)\end{tabular}}} \\\cline{2-9}
\multicolumn{1}{c}{}                                 & \textbf{FPE}                               & \textbf{FNE}                              & \textbf{FPE}                             & \textbf{FNE}                             & \textbf{FPE}                                 & \textbf{FNE}                                 & \textbf{FPE}                               & \textbf{FNE}                              \\\midrule
MEDPSeg (ours)                                       & $0.1\pm0.04$                               & $0.13\pm0.06$                             & $0.14\pm0.08$                            & $0.06\pm0.02$                            & $0.19\pm0.08$                                & $0.17\pm0.12$                                & $0.02\pm0.01$                              & $0.02\pm0.01$                             \\\midrule
Vessel 3D nnUNet                                     & $0.08\pm0.05$                              & $0.14\pm0.08$                             & N/A                                      & N/A                                      & N/A                                          & N/A                                          & N/A                                        & N/A                                       \\\midrule
Airway 3D nnUNet                                     & N/A                                        & N/A                                       & $0.07\pm0.06$                            & $0.07\pm0.04$                            & N/A                                          & N/A                                          & N/A                                        & N/A                                       \\
Zheng et al.                                         & N/A                                        & N/A                                       & $0.13\pm0.24$                            & $0.17\pm0.23$                            & N/A                                          & N/A                                          & N/A                                        & N/A                                       \\\midrule
Biondi et al.                                        & N/A                                        & N/A                                       & N/A                                      & N/A                                      & $0.24\pm0.13$                                & $0.49\pm0.14$                                & N/A                                        & N/A                                       \\
AMFM                                                 & N/A                                        & N/A                                       & N/A                                      & N/A                                      & $0.72\pm0.13$                                & $0.2\pm0.09$                                 & N/A                                        & N/A                                       \\
COPLE-Net                                            & N/A                                        & N/A                                       & N/A                                      & N/A                                      & $0.15\pm0.06$                                & $0.19\pm0.14$                                & N/A                                        & N/A                                       \\
Lesion 3D nnUNet                                     & N/A                                        & N/A                                       & N/A                                      & N/A                                      & $0.29\pm0.12$                                & $0.09\pm0.1$                                 & N/A                                        & N/A                                       \\\midrule
Lungmask                                             & N/A                                        & N/A                                       & N/A                                      & N/A                                      & N/A                                          & N/A                                          & $0.03\pm0.05$                              & $0.03\pm0.01$                                 \\\bottomrule                 
\end{tabular}
}
\caption{Comparing MEDPSeg's multitasking results in false positive and false negative error (FPE and FNE) in different test data from each target format. Lower is better.}
\label{tab:fpefne}
\end{table*}
AMFM and Biondi et al. are based on traditional feature engineering machine learning and have significantly lower lesion segmentation performance than MEDPSeg. There is no statistically significant Dice difference between MEDPSeg and the compared deep learning-based methods specialized in Lesion and Vessel segmentation. Note that the test split of the Lesion cohort is very challenging to all methods due to the inclusion of three different datasets: IdeiaGov, MOSMED, and MICCAI challenge. Regarding whole lung parenchyma segmentation performance, built in a polymorphic way through the union of healthy, GGO and consolidation outputs, MEDPSeg performed significantly better in lung parenchyma segmentation of severe pneumonia patients than Lungmask~\citep{hofmanninger2020automatic}. When analyzing false positive error (FPE) and (FNE), MEDPSeg is never worse in both when compared to the specialized methodologies (Tab~\ref{tab:fpefne}).

Finally, when looking at the proper largest connected tubular component segmentation metrics, MEDPSeg performs comparably to specialized nnUNets in the pulmonary artery and airway targets. Note that Zheng et al.‘s method was not trained in the ATM22 challenge data, while the specialized nnUNets are trained on the same development data. (Tab~\ref{tab:topology}). 

\begin{table}[ht]
\centering
\resizebox{\columnwidth}{!}{
\begin{tabular}{cccc}
\toprule
\textbf{Airway}           & \textbf{MEDPSeg} & \textbf{Airway 3D nnUNet} & \textbf{Zheng et al.} \\\midrule
\textbf{BD}               & $84.15\pm8.14$   & $84.16\pm8.77$            & $71.59\pm15.89$       \\
\textbf{TD}               & $91.45\pm3.92$   & $90.82\pm5.19$            & $83.63\pm9.12$        \\\midrule
\textbf{Pulmonary Artery} & \textbf{MEDPSeg} & \textbf{Vessel 3D nnUNet} &  -                     \\\midrule
\textbf{BD}               & $82.47\pm9.32$   & $83.24\pm12.45$           &  -                     \\
\textbf{TD}               & $90.78\pm4.25$   & $91.31\pm5.96$            &  -              \\\bottomrule      
\end{tabular}
}
\caption{\label{tab:topology}Test performance on topological metrics tree length detected rate (TD) and branch detected rate (BD) on the airway and pulmonary artery segmentation tasks.}
\end{table}



In general, when compared to other methods, MEDPSeg simultaneously segments multiple pulmonary targets with comparable or better performance to specialized state-of-the-art methods. It achieves superior performance compared to the current state-of-the-art for the segmentation of consolidation and lung parenchyma with severe pneumonia. MEDPSeg can provide all 5 different involved labels with a low resource usage footprint and robust generalization capabilities both in low and high-resolution scans due to its 2.5D nature and its training data variability.

\subsection{Reproducibility and Computational Efficiency}
\label{sec:rep}
An important parallel consideration of MEDPSeg's development was keeping the method lightweight and providing a reproducibility tool as a python pip package, with a graphical user interface, designed for ease of use by both experts and non-experts in the medical or engineering fields. Open source code and trained weights for reproducing our segmentation capabilities are in \url{https://github.com/MICLab-Unicamp/medpseg}. The only requirement for usage of the tool is a Python pip or conda environment. Due to working in a slice-wise manner, MEDPSeg can run in relatively low computer specifications for fast performance of less than 1 minute per high-resolution scan. Usage of the tool can be summarized as selecting a folder or CT scan file and selecting the output location for all segmentation masks, volume statistics, and lung involvement reports separated by lung and lobe (an external lung lobe segmentation method is also included). In addition, the tool supports upload of a single axial slice, for quick assessment of specific slices. Our repository also allows for complete reproducibility of our results in the CoronaCases and SemiSeg datasets, with the provided script reproducing all pipeline steps, from downloading the original data to outputting a results table. Due to the efficient design originating from the use of EfficientNet and EfficientDet, MEDPSeg can bring better performance while needing less multiplication and additions (MACS) than other architectures, scaling from 1 target to 5 very efficiently due to the way the segmentation heads share BiFPN and backbone weights (Tab.~\ref{tab:efficiency}).

\begin{table}[ht]
\centering
\resizebox{\columnwidth}{!}{
\begin{tabular}{lllll}
\toprule
\textbf{Method} & \textbf{MACS (G)} & \textbf{Fw/Bw size (GB)} & \textbf{N. Params. (M)} & \textbf{N. Targets} \\ \midrule
S-MEDSeg        & 28                & 6                                   & 62                         & 1                   \\ 
MEDPSeg         & 73                & 14                                  & 64                         & 5                   \\ 
AttUNet         & 184               & 3                                   & 14                         & 1                   \\ 
PolyAttUNet     & 325               & 5                                   & 23                         & 5                   \\ \bottomrule
\end{tabular}
}
\caption{Multiply-add count (MACS), size in memory of forward and backward computing (Fw/Bw size), numbers of parameters (N. Params.) and number of targets (N. Targets) processed by MEDPSeg, PolyAttUNet and their target specialized variations.}
\label{tab:efficiency}
\end{table}

MEDPSEg performs segmentation of all targets simultaneously in a single forward pass of one network and does not require any post-processing other than stacking axial predictions. As a result, MEDPSeg is able to provide a complete high-resolution 3D pulmonary segmentation of 6 different targets in less than 1 minute with a NVidia 2060 GPU. Even without considering the logistical difficulty of training and using separate specialized methods for each target, one can achieve five times faster prediction on average with MEDPSeg in comparison to, for example, performing inference with five separately trained nnUNet.

\subsection{Qualitative Results}
\label{sec:quali}
\begin{figure}[ht]
\centerline{\includegraphics[width=\columnwidth]{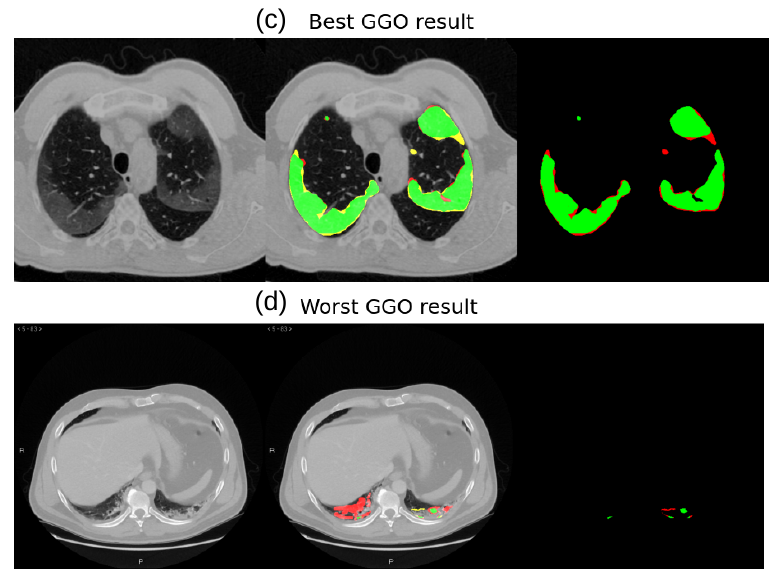}}
\caption{\label{fig:ggo_bnwsemiseg}The best test case for GGO segmentation from the SemiSeg dataset with 0.92 Dice and the worst with 0.13 Dice. Consolidation annotation is omitted. Green denotes overlap, red denotes false negatives, and yellow denotes false positives. }
\end{figure}

\begin{figure}[ht]
\centerline{\includegraphics[width=\columnwidth]{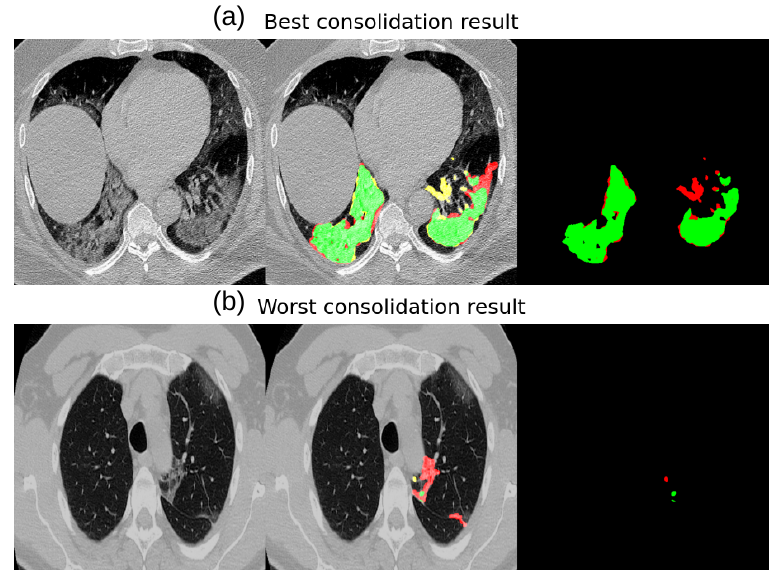}}
\caption{\label{fig:con_bnwsemiseg}The best test case for consolidation segmentation from the SemiSeg dataset with 0.85 Dice, and the worst with 0.08 Dice. GGO annotation is omitted. Green denotes overlap, red denotes false negatives, and yellow denotes false positives.}
\end{figure}

For GGO and consolidation 2D segmentation in the SemiSeg dataset, best and worst results according to the manual ground truth are illustrated in Figures~\ref{fig:ggo_bnwsemiseg} and \ref{fig:con_bnwsemiseg}. As expected, the more difficult cases are those with small regions of GGO or consolidation. The subjective nature of opacity definition makes this problem difficult even for humans. 

\begin{figure*}[ht]
\centerline{\includegraphics[width=.85\textwidth]{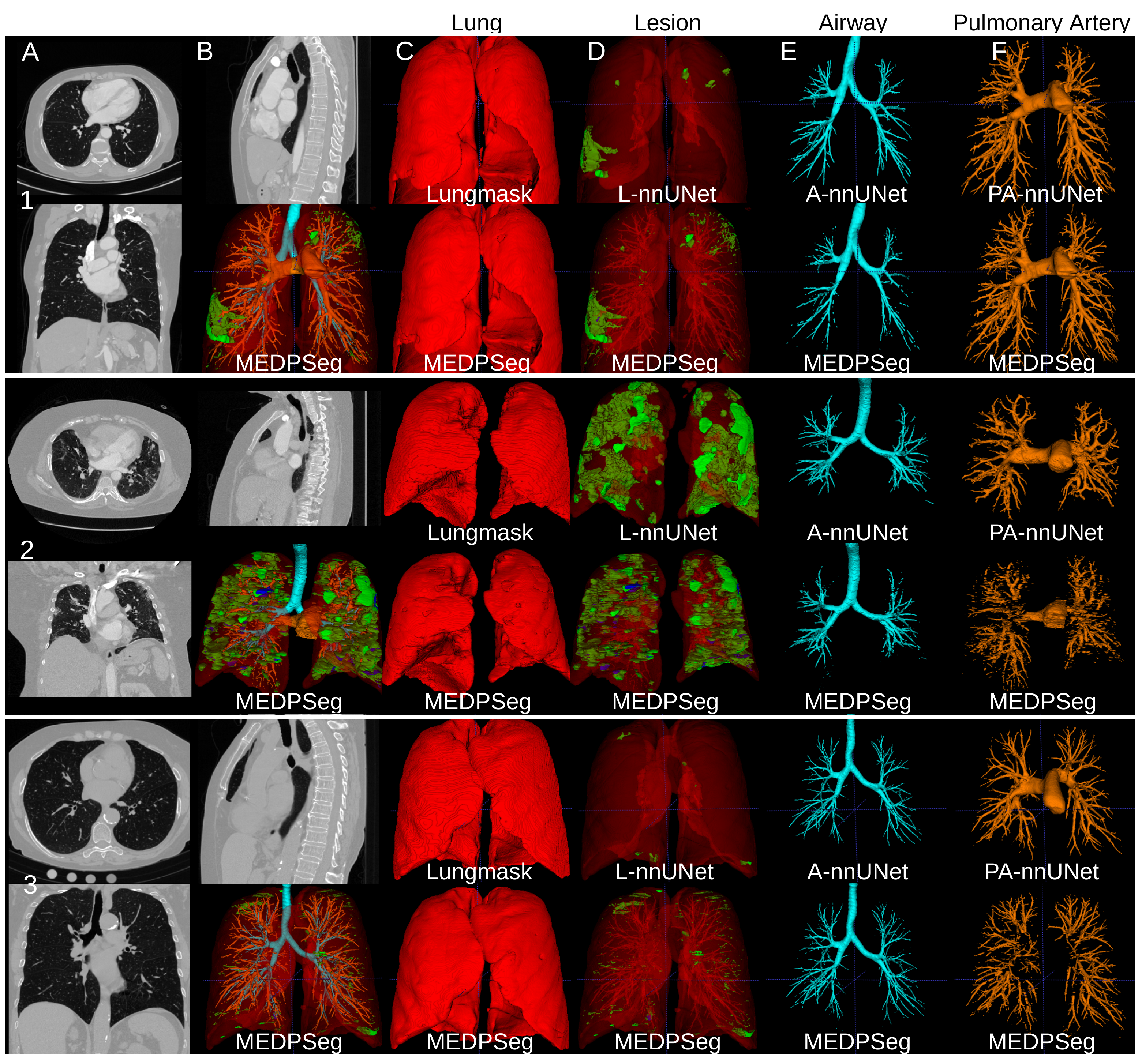}}
\caption{\label{fig:qualitative}(A) Mid-axial and coronal slice, (B) mid-sagittal slice and 3D rendering of MEDPSeg prediction. Each component of MEDPSeg's prediction is compared to other specialized methods, including: (C) Lungmask, (D) lesion trained nnUNet, (E) airway trained nnUNet, and (F) vessel trained nnUNet. Each numbered row represents a different scan from a different dataset not included in annotated data. Data sources: (1) PARSE, (2) COVID-19 NY-SBU, (3) COPD-TLC.}
\end{figure*}

Following we examine the multitasking performance of MEDPSeg in unseen scans from silver standard data, when compared to specialized methods Lungmask, L-nnUNet (Lesion), A-nnUNet (Airway), and P-nnUNet (Pulmonary artery), the same nnUNet used in quantitative comparisons. Figure~\ref{fig:qualitative} shows raw axial and coronal slices and 3D renderings of results from MEDPSeg and state-of-the-art methods in multiple challenging external cases, from CT scans not included in manually annotated data nor development or testing of MEDPSeg. There are scans of high lung involvement due to COVID-19, changing field of view, small lung volume due to lung lesion, lower resolution, or no contrast for pulmonary artery detection. MEDPSeg will multitask all pulmonary targets without signs of overfitting, a low resource usage footprint, and robust generalization capabilities due to its 2.5D nature and its training data variability.

MEDPSeg demonstrates impressive adaptability to heterogeneous unseen scans from multiple sources, fields of view, resolution, and other parameters, producing results similar to specialized methods from each target. Limitations are noticeable in attempting to segment tubular tree-like structures in low-resolution scans. In addition, performance of pulmonary artery segmentation is degraded in non-contrast scans. These limitations are also observed by human raters when generating manual annotations~\citep{luo2023efficient}.


\section{Discussion}
\label{sec:discussion}

Even with the low availability of gold standard for GGO and consolidation annotations, MEDPSeg was able to learn from an enormous amount of data during training, using the proposed hierarchical polymorphic multitask learning strategy to leverage heterogeneous targets and the hierarchical nature of lung lesion annotation, achieving state-of-the-art results in the separation of lung lesion into GGO and consolidation. HPML allows one to mix target data of different formats from different partially labeled datasets, subsequently learning from these datasets. A similar methodology could be applied to other problems which have heterogeneous labels available, from hierarchical structures with varying degrees of specificity in annotation. This strategy results in more data variability in training and allows for indirect optimization of complex targets while leveraging more readily available simpler targets. We have demonstrated performance improvements in the task of lesion segmentation and separation into GGO and consolidation by using HPML both in our proposed architecture and in a modification of a UNet-like architecture.

MEDPSeg's polymorphic multitasking results segmenting lung, lung lesion, pulmonary artery, and airway do not only confirm our initial hypothesis and increase GGO and consolidation segmentation performance, but are statistically similar or significantly better in overlap metrics than multiple state-of-the-art methods trained specifically for each involved target. Of note is the statistically significant improvement in consolidation separation and lung parenchyma with high opacity involvement. Overall, these results confirm the importance of data variability and neighboring anatomy guidance during the training of medical imaging segmentation models, and more so, present the impact of label variety. In addition, silver standard pertaining (SSP) was the best strategy for the employment of silver standard data. This most likely is due to increased data shown to the network, with the silver standard lung label in further training avoiding the forgetfulness effect~\citep{cossu2022continual} regarding pretraining.

As far as we know, MEDPSeg is the first method to provide the segmentation of all the involved pulmonary targets, with the added benefit of an open-source software tool implementation for the reproducibility of its lung assessment capabilities and evaluation.

\subsection{Limitations}
MEDPSeg is not without limitations. MEDPSeg will still try to predict vessels and airways in scans with high distance between slices but results will not be correct due to these structures not being visible in low resolution CT. Moreover, gold standard training samples for pulmonary artery segmentation were contrast-enhanced CTs, meaning in scans without contrast pulmonary artery results outside of the lung will be of lower quality due to low mediastinal contrast. In GGO and consolidation annotations, very small areas can be missed (Fig.~\ref{fig:qualitative}), a phenomenon that is common to all competing methods currently as shown by the high standard deviation in Dice (Tab.~\ref{tab:semiseg_compare}). Another problem that is still open is that of the highly variable annotation protocols between lesion datasets, clear in the high standard deviation of the Lesion cohort test results, for all methods (Fig.~\ref{fig:boxplots}. Finally, the implementation of the training process is difficult regarding data preparation, balancing of the batch, and the computation of loss for each target format, requiring custom implementation for each problem. Moreover, we use the same data preprocessing and augmentation for all target formats. Custom strategies for different targets could improve results. Nevertheless, the reproducibility of the prediction capabilities of trained MEDPSeg is easily available in our online repository, and future work will explore the automation of PML through decoder conditioning~\citep{zhang2021dodnet} to any problem that could benefit from hierarchical labels and multitasking, removing the need for custom implementations.




\section{Conclusion}
\label{sec:conclusion}

This work proposes MEDPSeg, a new lightweight methodology that leverages information from multiple partially labeled datasets with conflicting manual and automated target formats, using hierarchical polymorphic multitask learning to achieve state-of-the-art performance in multiple chest CT segmentation tasks, especially in the segmentation of GGO and consolidation. We showcase how one could use the hierarchical properties of the lung anatomy and its internal lesions to indirectly optimize the segmentation of detailed targets while mixing in simpler and more readily available general lesion and lung annotation. Moreover, benefits of multitasking circulatory and airway targets are demonstrated. Finally, we provide open-source trained weights and easy-to-use graphical tool for the reproducibility of MEDPSeg's evaluation and segmentation capabilities. We hope for MEDPSeg to be used to reliably characterize large chest CT cohorts in future studies.

\section*{Acknowledgements}

D. Carmo is supported by Sao Paulo Research Foundation (FAPESP) grants \#2019/21964-4 and \#2022/02344-8. R Lotufo and L Rittner
are partially supported by CNPq (The Brazilian National Council for
Scientific and Technological Development) under grants 313047/2022-7
and 313598/2020-7, and Coordenação de Aperfeiçoamento de Pessoal de Nível Superior (CAPES) under grant 506728/2020-00.

\bibliographystyle{model2-names.bst}\biboptions{authoryear}
\bibliography{main}

\end{document}